# Spin scattering of a particle for periodic boundary conditions

L.L. Chu *, K.W. Yu

*Department of Physics, The Chinese University of Hong Kong, Shatin, New Territories, Hong Kong, China*

**Abstract**

We have studied anomalous diffusion of a particle in a random medium in which the passage of the particle may modify the state of the visited sites. The simplicity of the dynamics allows analytic solution. Interesting propagation and organization behaviors will be reported. © 2001 Elsevier Science B.V. All rights reserved.

*PACS:* 05.50.+q; 05.60.+w; 05.20.Dd; 82.40.C

*Keywords:* Spin scattering; Random media; Lattice models; Periodic boundary condition

## 1. Introduction

Anomalous diffusion of a particle in a random medium has been a well studied topic in condensed matter physics. The most striking behavior in the particle dynamics is observed when the passage of the particle modifies the state of the visited sites. The investigation can naturally be extended to many-particle systems.

Boon and his coworkers [1] considered a prototype in which a particle is moving on a one-dimensional lattice whose sites are occupied by scatterers with the following properties:
 (i) the state of each site is defined by its spin (up or down);
 (ii) the particle arriving at a site is scattered forward (backward) if the spin is up (down);
(iii) the state of the site is modified by the passage of the particle (up ⇔ down).

A simple case is shown in Fig. 1(a). At time $t = 0$, the spin of the first site is down while the spins at other sites are randomly distributed on the lattice with a probability $p$ of up spins. The particle is initially located at the first lattice site, and its velocity $v$ is one step per unit of time to the right. They found that after the passage of the particle through the one-dimensional lattice, all the spins are in a state opposite to their initial state (↑ ⇔ ↓) with one lattice site shift in the direction opposite to the propagation direction. They also found that the average propagation velocity is given by

$$\langle v \rangle = \frac{1}{3 - 2p}. \qquad (1)$$

Moreover, they showed that the particle moving on a one-dimensional lattice will visit a site with spin down (initially) twice and a site with spin up (initially) once, i.e. it spends 2 more units of time on passing a lattice site with spin down. Hence, the propagation velocity $v$ is given by

$$v = \frac{N_{\text{up}} + N_{\text{down}}}{N_{\text{up}} + 3N_{\text{down}}}, \qquad (2)$$

* Corresponding author.
 *E-mail address:* llchu@phy.cuhk.edu.hk (L.L. Chu).

(a)

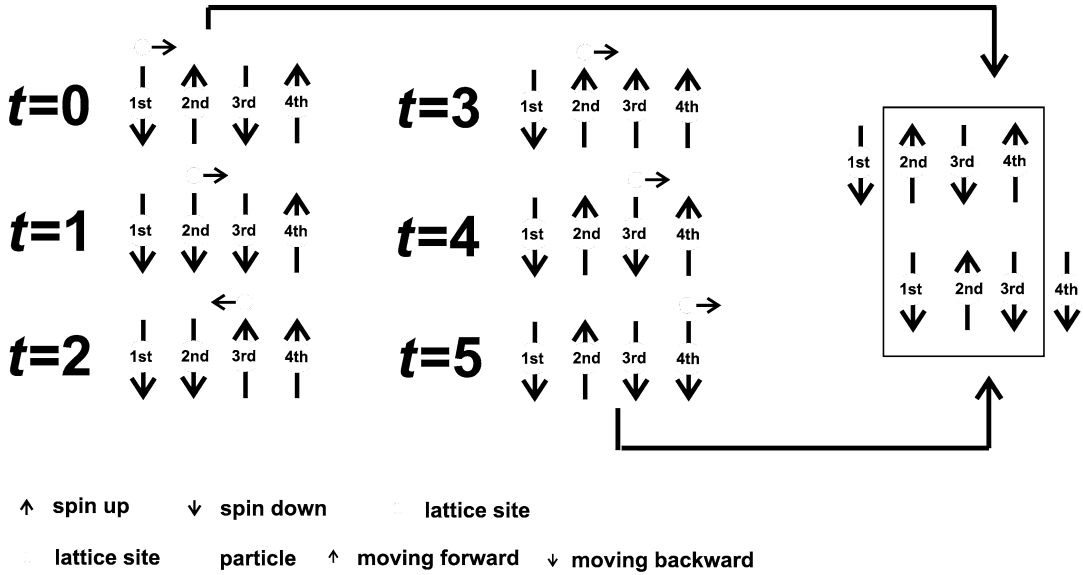

(b)

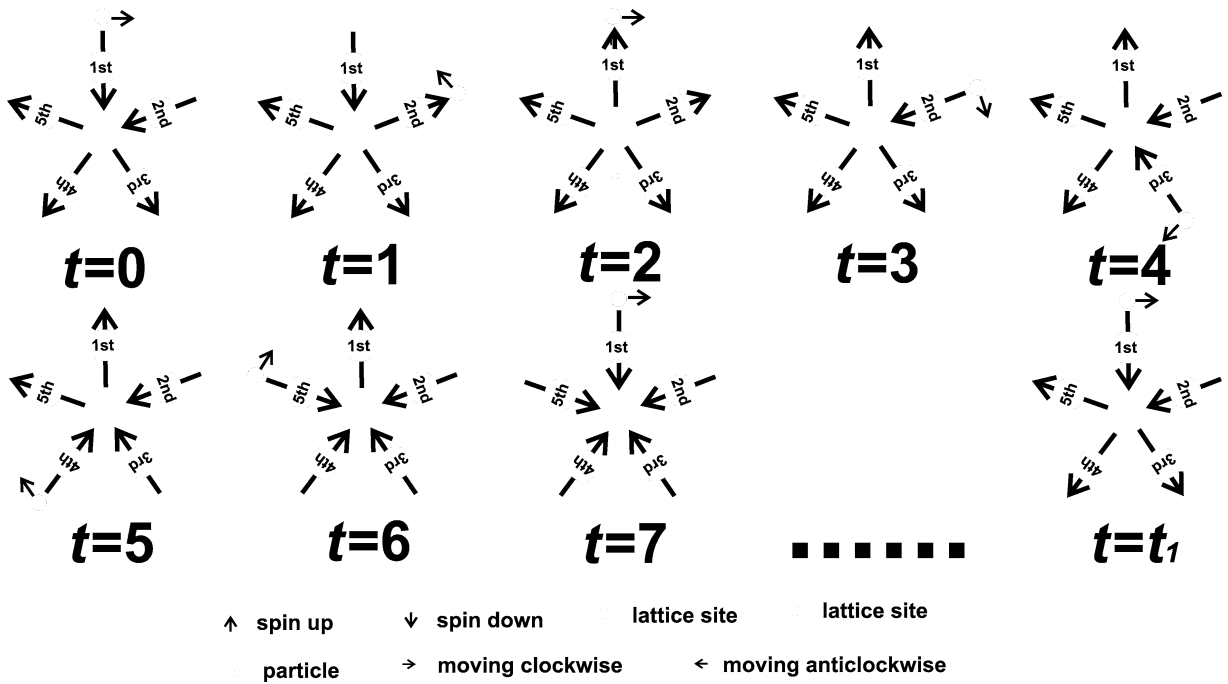

Fig. 1. (a) Propagation of a single particle on a one-dimensional lattice; (b) propagation of a single particle on a circular lattice.

where $N_{up}$ and $N_{down}$ are the numbers of sites (visited by the particle) with spin up and spin down, respectively.

## 2. Periodic boundary condition

By extending the above ideas to the periodic boundary condition, i.e. by identifying the first and last sites of the chain, Boon and his coworkers proposed that if the particle propagates clockwise on the circle, there will be a counterclockwise drift (two lattice sites back) of the initial spin configuration every two cycles, and the average velocity will be given by:

$$\langle v \rangle = \frac{1}{2}\left( \frac{1}{3 - 2p} + \frac{1}{3 - 2(1 - p)} \right). \quad (3)$$

We perform numerical simulation in which a single particle moves on a circular lattice as shown in Fig. 1(b). The spin configuration and the propagation rules of the particle remain the same as those proposed by Boon. At time $t = 0$, the spin of the first site is down while the spins at other sites are randomly selected as either up or down. The particle is located at the first site, and its velocity $v$ is one step per unit of time in the clockwise direction. At $t = 7$ the particle goes back to its initial position with a clockwise propagation velocity $v$, we say it has just finished one cycle on the lattice. At $t = t_1$ the particle has just finished one cycle and the spin configuration is the same as the initial one, we say the spin configuration has just undergone a change of one period and $T = t_1$.

## 3. Spin configuration

Now we define "1" and "0" as spin up ($\uparrow$) and down ($\downarrow$), respectively. Consider two examples shown in Fig. 2. We use "00111" to represent the initial configuration of the lattice in Fig. 1(b) and consider one more lattice with spin configuration "01111". The difference between the initial spin configurations of the two lattices is the spin state of the second site.

We find that the shift pattern of the spin is not exactly the same as that proposed by Boon. Instead, at the end of one cycle the spin configuration of the lattice has the following characteristics:

 (i) the first site is spin down;
 (ii) the spin state of the last site is the same as that of the second site in the previous cycle;
(iii) all the remaining spins are in the state opposite to their state in the previous cycle with one lattice site shift in the direction opposite to the propagation direction.

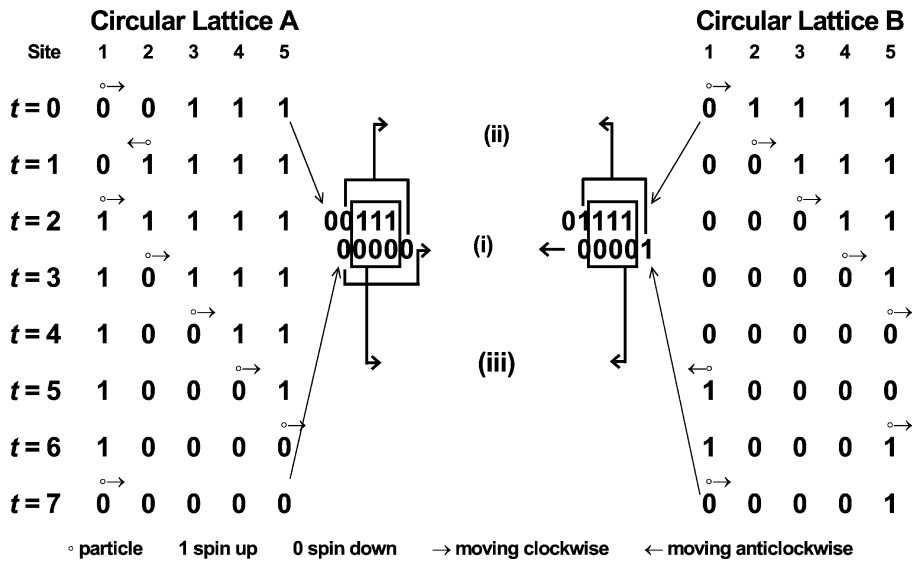

Fig. 2. Change of spin configurations of two circular lattices within a cycle.

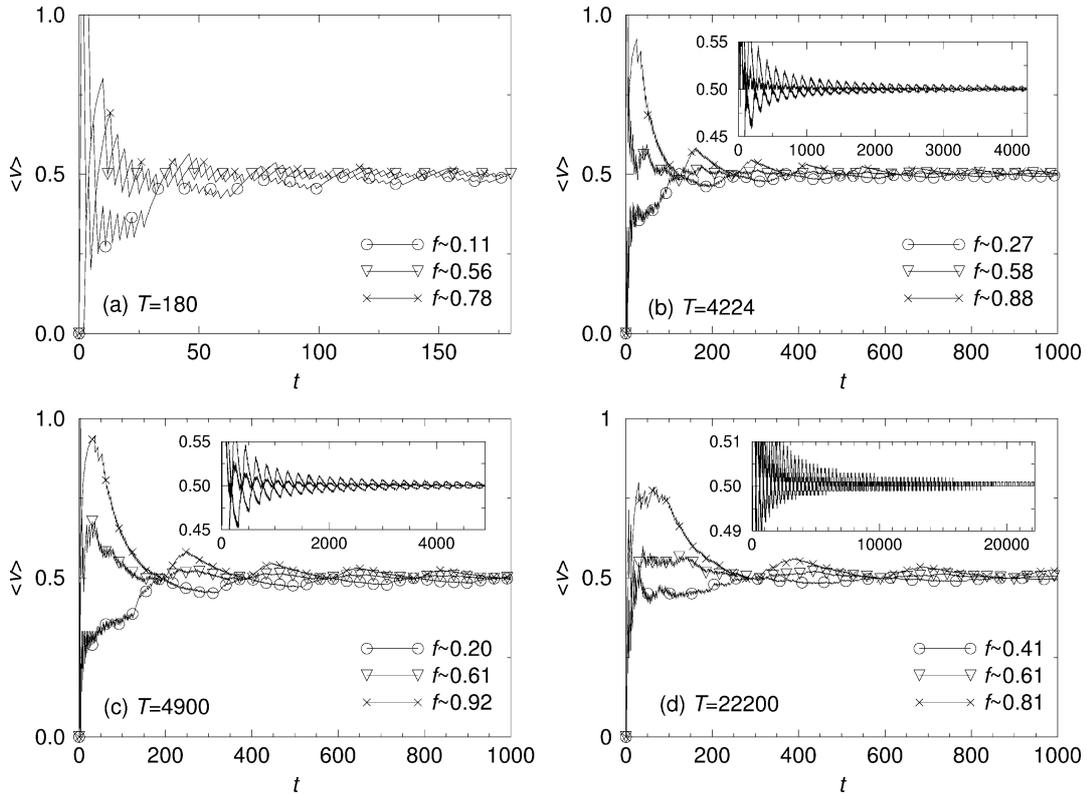

Fig. 3. Average velocity of the particle against time for several circular lattices.

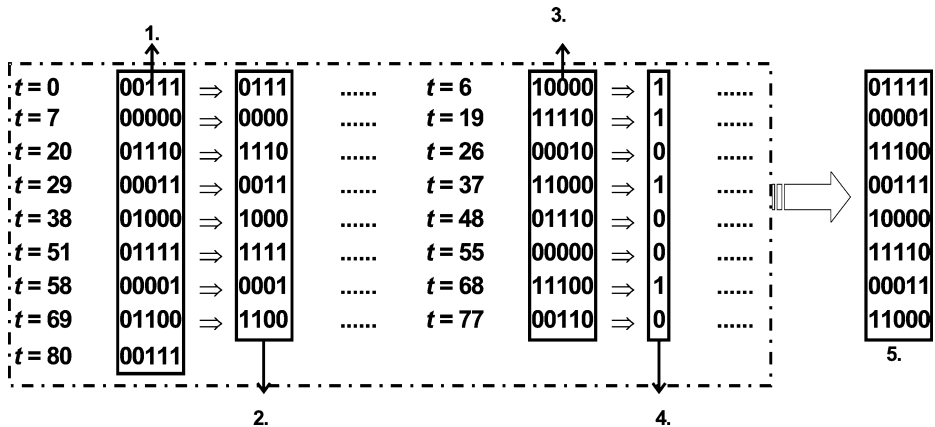

Fig. 4. Conversion of a circular lattice to its effective one-dimensional lattice.

## 4. Average propagation velocity

Our result shows that the average propagation velocity is given by $\langle v \rangle = 1/2$, regardless of the initial spin configuration. We also find that the period of the spin configuration $T$ is finite. The average propagation velocity $\langle v \rangle$ fluctuates around $1/2$ and approaches exactly $1/2$ at the end of each period. Several examples are shown in Fig. 3. The spin up fraction of the lattice (excluding the first site) is denoted by $f$.

The striking results can be explained by the fact that a particle moving on a circular lattice can be regarded as a particle moving on an effective one-dimensional lattice. An example is shown in Fig. 4: column 1 shows the spin configuration at the end of each cycle, i.e. the particle arrives at the first lattice site with a clockwise velocity; column 2 shows the spin states of the next four lattice sites, i.e. the second to last lattice sites seen by the particle, at that moment; column 3 shows the spin configuration when the particle arrives at the last site with a clockwise velocity at each cycle; column 4 shows the spin state of the next lattice site, i.e. the first lattice seen by the particle, at that moment. According to the information given by columns 2 and 4, in the moving frame of the particle, it propagates on a one-dimensional lattice with an initial spin configuration shown in column 5 which consists of half spin up lattice sites (excluding the first lattice site). Fig. 5 shows the position of the particle on this circular lattice and its effective one-dimensional lattice against time. After the passage of a particle through a circular lattice till the spin pattern undergoes one period's change, it has visited equal numbers of sites with spin up and spin down. According to Eq. (2), the propagation velocity $v$ of the particle is therefore equal to 1/2.

## 5. Conclusions

In summary, when a single particle moves on a periodic boundary condition, its average propagation velocity $\langle v \rangle$ is equal to 1/2, regardless of the initial spin configuration because the situation is the same as that a particle moving on an effective one-dimensional lattice having equal numbers of sites with spin up and spin down (excluding the first lattice site). When the particle arrives at the first site of the lattice at the end of each cycle, the spin configuration has the following characteristics:

(i) the first site is spin down;
(ii) the spin state of the last site is the same as that of the second site of the previous cycle;
(iii) all the remaining spins are in the state opposite to their state in the previous cycle with one lattice site shift in the direction opposite to the propagation direction.

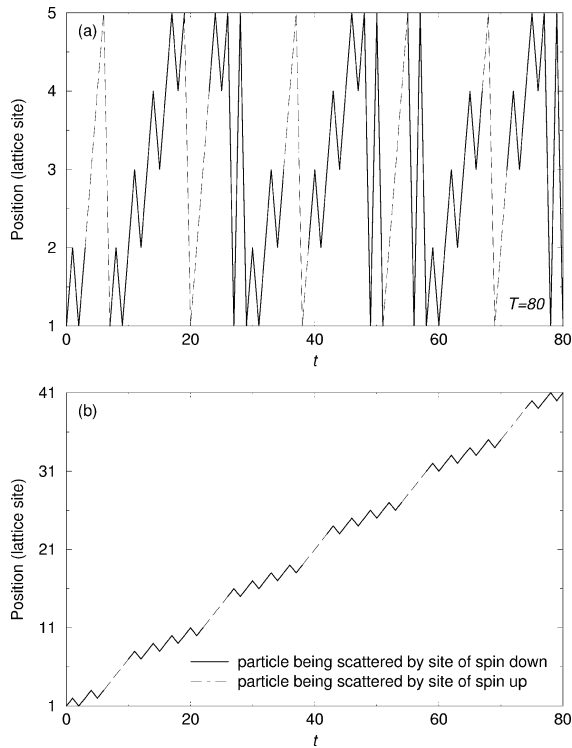

Fig. 5. (a) Position of particle on circular lattice (with 00111 as initial configuration) against time. (b) Position of a particle on (a)'s effective one-dimensional lattice against time.

## Acknowledgements

L.L. Chu gratefully acknowledges financial support from the Graduate School of the Chinese University of Hong Kong for Overseas Academic Activities, and from Department of Physics of the Chinese University of Hong Kong for providing the Academic Exchange Fund.